\documentclass[twocolumn,superscriptaddress,showpacs,preprintnumbers,amsmath,amssymb,prb,floatfix]{revtex4}

\usepackage{graphicx}
\usepackage{dcolumn}
\usepackage{bm}
\newcommand{\bra}[1]{\left\langle #1 \right|}
\newcommand{\ket}[1]{\left| #1 \right\rangle}

\newcommand{\abinitio}[0]{\textit{ab~initio}}

\newcommand{\bucky}[0]{C$_{60}$}
\newcommand{\sbucky}[0]{C$_{28}$}
\newcommand{\hsbucky}[0]{C$_{28}$H$_4$}
\newcommand{\etal}[0]{\textit{et al.}}
\newcommand{\ie}[0]{i.e.}
\newcommand{\eg}[0]{e.g.}
\newcommand{\Tc}[0]{$T_c$}

\newcommand{\Figure}[0]{Fig.}
\newcommand{\Equation}[1]{Eq.~(#1)}

\begin{document}

\title{Electron-Phonon Interactions in C$_{28}$-derived Molecular Solids}

\author{Nichols A. Romero}
\affiliation{Department of Physics, Materials Research Laboratory, and Materials Computation Center, University of Illinois, Urbana, IL 61801}
\author{Jeongnim Kim}
\affiliation{NCSA, Materials Research Laboratory, and Materials Computation Center, University of Illinois, Urbana, IL 61801}
\author{Richard M. Martin}
\affiliation{Department of Physics, Materials Research Laboratory, and Materials Computation Center, University of Illinois, Urbana, IL 61801}

\date{August 27, 2004}

\begin{abstract}
We present {\it ab initio} density-functional calculations of
molecular solids formed from \sbucky-derived closed-shell
fullerenes. Solid \hsbucky{} is found to bind weakly and exhibits many
of the electronic structure features of solid \bucky{} with an
enhanced electron-phonon interaction potential. We show that chemical
doping of this structure is feasible, albeit more restrictive than its
\bucky~counterpart, with an estimated superconducting transition
temperature exceeding those of the alkali-doped \bucky{} solids.
\end{abstract}


\maketitle

There continues to be intense research efforts in studying fullerenes
for their diverse properties, not the least of which is their
unusually high superconducting transition temperatures \Tc{} (up to 40
K in Cs$_3$\bucky{}).\cite{ttmpalstra}. Theoretical studies show that
many of the phenomena associated with this class of materials can be
explained within the electron-phonon mediated picture of
superconductivity.\cite{cmvarma,mschluter1,iimazin,vpantropov,ogunnarsson1}
The high \Tc~of these materials relative to that of intercalated
graphite is attributed to the curvature of
\bucky{}.\cite{mschluter1,jlmartins2,vhcrespi} Hence, solids based on
smaller fullerenes possessing an electronic structure similar to that
of alkali-doped \bucky{} may exhibit a \Tc{} enhancement.

The goal of the present work is to identify new molecular solids
analogous to the alkali-doped \bucky{} solids with similar electronic
structure and increased electron-phonon coupling.  We consider solids
composed of C$_{28}$-derived closed-shell molecules (C$_{24}$B$_4$,
C$_{24}$N$_4$, and C$_{28}$H$_4$) which can potentially have a large
tunable density-of-states (DOS) arising from the narrow weakly
broadened bands. Theoretical studies of the isolated molecules
indicate that they should be
stable,\cite{tguo2,ekaxiras2,dfttuan,ynmakurin} however, to our
knowledge there are no studies of their solid forms.  Here we present
results of \abinitio{} pseudopotential density-functional calculations
to determine the structural and electronic properties of the solids in
pristine and doped forms.  There have been \abinitio{} studies of
solids based on~\sbucky{}
(Refs.~\onlinecite{dmbylander,ekaxiras1,nbreda}) and
C$_{36}$,\cite{mcote,jcgrossman} which exhibit increased
electron-phonon coupling. However, these molecules form covalent
solids differently from \bucky{}.  We find that C$_{24}$B$_4$ and
C$_{24}$N$_4$ also form strongly-bonded solids, but that \hsbucky{}
exhibits many of the salient features of solid \bucky{}. Several
doping scenarios are investigated for alkali doped \hsbucky{}
crystals, and a promising candidate is identified for high \Tc{}.

First proposed by Kroto,\cite{hkroto} the \sbucky{} molecule obeys
the $T_d$ point group. This fullerene is produced with an abundance
nearly as great as that of \bucky{} in the laser vaporization of
graphite.\cite{tguo1} It is a very reactive molecule with dangling
bonds localized on each of the four apex atoms shown in black for each
\sbucky{} in \Figure~1(a). Because the \sbucky{} molecule is a
tetravalent superatom, a reasonable candidate for a solid is a
four-fold coordinated diamond-like structure. The \sbucky{} molecules
can be four-fold coordinated with respect to the tetrahedron apexes as
in \Figure~1(a) \textit{apex-bonded} or with respect to the
tetrahedron faces as in \Figure~1(b) \textit{face-bonded}. Since the
\sbucky{} molecules have dangling bonds localized on the tetrahedron
apexes, the lower energy structure will be the former. Previous
\abinitio{} calculations \cite{dmbylander,ekaxiras1} have shown that
the covalent bonds of the apex-bonded \sbucky{} hyperdiamond are very
strong compared to the weak forces responsible for bonding in solid
\bucky{}.

\begin{figure}[htb]
\centering \includegraphics[width=3.3in]{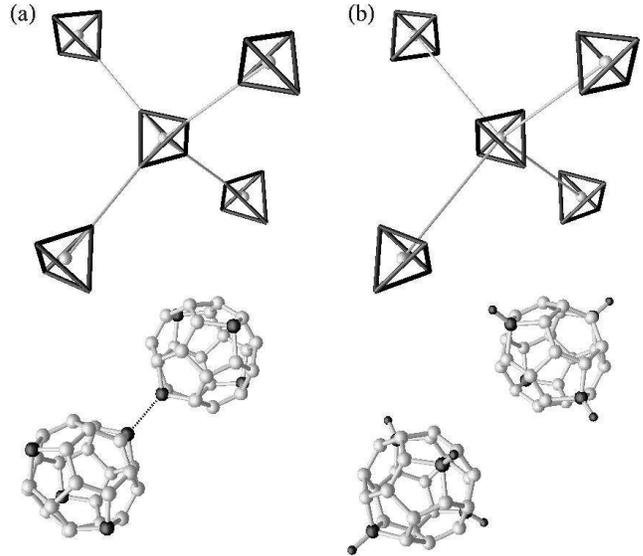}
\caption{\label{structure} Schematic diagrams of the first nearest
neighbors in the hyperdiamond structures (top) with examples of pairs
of \sbucky-derived molecules shown below. The lattice sites in the
solid are highlighted by the grey spheres enclosed in the
tetrahedra. The two distinct orientations of the constituent molecules
are represented by those of the tetrahedra. The nearest-neighbor pairs
show two distinct bonding configurations: (a)~\textit{apex-bonded}
\sbucky{} hyperdiamond forms covalent bonds (dotted line) between apex
atoms colored in black; (b)~\textit{face-bonded} \hsbucky{}
hyperdiamond forms weak bonds between six-membered rings with hydrogen
atoms depicted as smaller black spheres.}
\end{figure}

In search of suitable molecular solids, we study hyperdiamond
structures constructed from three distinct closed-shell molecules --
C$_{24}$B$_4$, C$_{24}$N$_4$, and C$_{28}$H$_4$, using \abinitio{}
pseudopotential density functional methods within the local density
approximation (LDA).  The SIESTA code \cite{pordejon,jmsoler} was used
to perform conjugate gradient minimization of forces and stresses to
find relaxed structures. Though other possible solid structures for
\sbucky{} have been proposed,\cite{jkim} they are reported to be
higher in energy than apex-bonded hyperdiamond and so were not
considered in this study. Although all three molecules are stable, as
was found previously, \cite{tguo2,ekaxiras2,dfttuan,ynmakurin} we find
that in the solid the C$_{24}$B$_4$ molecules are unstable and break
apart.  Of the two geometries investigated for C$_{24}$N$_4$,
apex-bonded hyperdiamond is much lower in energy forming a covalent
solid like \sbucky{}. The face-bonded C$_{24}$N$_4$ hyperdiamond is
higher in energy and in addition shows significant hybridization in
the conduction band. Thus neither C$_{24}$B$_4$ nor C$_{24}$N$_4$ were
suitable candidates for reproducing the electronic structure of solid
\bucky{}.

The \hsbucky{} molecule is found to be stable in the solid.  Because
the dangling bonds are passivated by hydrogen, apex-bonded \hsbucky{}
hyperdiamond (and other suggested structures~\cite{jkim}) are not
favorable; the lower energy structure is the face-bonded \hsbucky{}
hyperdiamond shown in \Figure~1(b).  The latter is predicted to be a
weakly bound solid with with a lattice constant of 16.3~\AA{} and a
binding energy of approximately 0.2~eV per \hsbucky{} molecule. The
bonding between the six-membered rings is similar to that found for
some orientations of \bucky{} molecules.\cite{ogunnarsson2} The
structural and electronic properties of face-bonded \hsbucky{}
hyperdiamond in both its pristine and doped forms are the subject for
the remainder of this paper.

\begin{figure}[htb]
\centering \includegraphics[width=2.4in,angle=-90]{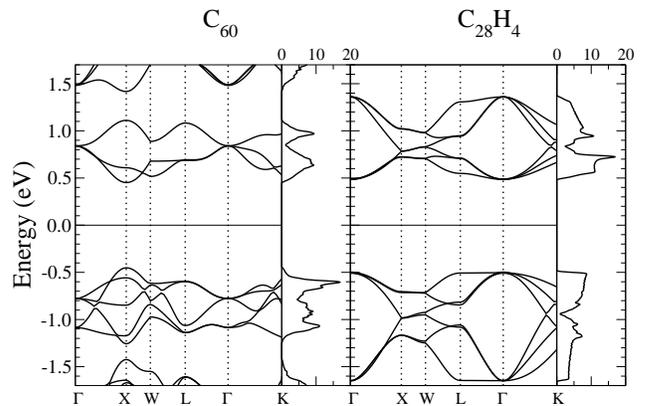}
\caption{\label{combo_bands_dos} Comparison of the band structure and
DOS (states/eV/spin/cell) between solid fcc-Fm$\bar{3}$ \bucky~and
face-bonded \hsbucky~hyperdiamond.  For \hsbucky{} solid, the set
of six bands above and below the gap are derived respectively from the
three-fold degenerate LUMO and HOMO for each \hsbucky{} in the
two-molecule cell. The thin solid lines at zero show the centers of
the direct gaps.}
\end{figure}

Figure~\ref{combo_bands_dos} compares the band structure and DOS
between solid \bucky{} and \hsbucky{} around the band gap. The valence
(conduction) band is formed from the three-fold degenerate LUMO (HOMO)
of the C$_{28}$H$_4$ molecule. The weakly broadened bands bear a
striking similarity to those found in solid \bucky{}. Undoped solid
\hsbucky{} forms an insulator with 1-eV direct gap at $\Gamma$. Upon
chemical doping, the Fermi energy is expected to fall within a DOS
peak with a value comparable to the alkali-doped \bucky{} materials.

The dimensionless electron-phonon coupling parameter $\lambda =
N(0)V_\mathrm{ep}$ depends on both the DOS at the Fermi energy $N(0)$
and the electron-phonon interaction potential $V_\mathrm{ep}$ which is
proportional to the curvature of the
fullerene.\cite{jlmartins2,mschluter1,vhcrespi} That \hsbucky{} is
more curved than \bucky{} suggests that higher transition temperatures
are obtainable. In general, $V_\mathrm{ep}$ can be calculated as a
double sum over the Fermi surface connecting states due to the phonon
deformation potential. For molecular solids like \bucky{} and
\hsbucky{}, the small dispersion of the electronic and phononic
spectra implies that the molecular states and intramolecular phonons
are an excellent approximation to those found in the
solid.\cite{cmvarma,mschluter1,vpantropov} Therefore, we compute
$V_\mathrm{ep}$ for solid \hsbucky{} by only considering the
intramolecular phonon coupling to the molecular states.

The electron-phonon interaction potential is evaluated for the
three-fold degenerate LUMO using the phonon frequencies $\omega_\nu$
and eigenvectors $\varepsilon_\nu$ for the isolated
\hsbucky~molecule. Since our fullerene is composed of multiple
species, it is convenient to include the mass of each species into the
phonon eigenvectors normalization, $\sum_i^{\mathcal{N}}
\varepsilon^{i\dag}_\nu \cdot \varepsilon^i_{\nu^\prime} M_i =
\delta_{\nu\nu^\prime}$, where $\mathcal{N}$~is the number of atoms in
the fullerene. The electron-phonon interaction potential can be
extracted from $\lambda$ \cite{drainer} and written in the form 
\begin{eqnarray}
V_{\mathrm{ep}} = \frac{1}{g^2}\sum_\nu \frac{1}{\omega_\nu^2}
\sum_{\alpha,\alpha^\prime}^g |\bra{\alpha}\sum_i^\mathcal{N}\varepsilon_\nu^i \cdot
\nabla_i V \ket{\alpha^\prime}|^2, \label{Eqn:Vep}
\end{eqnarray}
where $\alpha$ and $\alpha^\prime$ are the molecular states with
degeneracy $g$ for which the coupling is being evaluated. The matrix
element in \Equation{\ref{Eqn:Vep}} is evaluated by a
finite-difference approach where the deformation potential is
calculated within the frozen-phonon scheme. Our calculated value of
$V_{\mathrm{ep}}$ is 181 meV, which includes contributions from most
of the intramolecular phonons. It is observed in Raman-scattering
experiments \cite{sjduclos,tpichler} and predicted by theory
\cite{mschluter3,vpantropov} that the $A_g$ phonons are screened out
in alkali-doped \bucky{}. In the case of \hsbucky{}, the $A_g$ phonons
contribute 12 meV to $V_{\mathrm{ep}}$. Even taking into account the
screening of the $A_g$ phonons, the electron-phonon interaction
potential for \hsbucky{} is over twice as large as that of \bucky~(63
meV).\cite{mcote}

In \Figure~2 we see that analogous to \bucky{}, solid \hsbucky{} is an
insulator and can not superconduct unless doped. {\it Ab~initio}
calculations of alkali-doped \bucky{} have demonstrated that
intercalation of alkali atoms into the tetrahedral and octahedral
sites result in the donation of the alkali valence electrons to the
\bucky{} conduction
band.~\cite{jlmartins1,mschluter1,ssatpathy,wandreoni} This is
reflected in the band structure by a lack of hybridization of the
alkali states with the conduction band, so that the band structure of
the superconducting alkali-doped \bucky{} differs from that of
pristine \bucky{} primarily by a rigid shift in the Fermi energy. This
is viewed as the ideal doping case which we seek in the smaller
fullerene solids.

We have studied doping of solid \hsbucky{} with the alkali Na; other
alkali atoms are expected to behave similarly. The intercalation of Na
atoms into the solid \hsbucky{} was investigated in three different
scenarios:~(a) Na\hsbucky{} intercalation into the tetrahedral site;
(b) Na$_2$\hsbucky{} intercalation into the interstitial site between
the six-membered rings on nearest neighbor \hsbucky{} molecules; (c)
Na@\hsbucky{} endohedral doping. In each case, the Na-doped \hsbucky{}
structures were relaxed by conjugate gradient minimization of
forces and stresses. We discuss the effects of doping as they pertain
to the crystal structure, electronic properties and enthalpies of
reaction.\footnote{The enthalpy of reaction is referenced to the
relevant constituents, the calculated total energy of solid \hsbucky{}
and bcc sodium metal, unless otherwise noted. Negative values for the
enthalpy indicate that the compounds are stable with respect to the
reference system.}

\begin{figure}[htb]
\centering \includegraphics[width=2.4in,angle=-90]{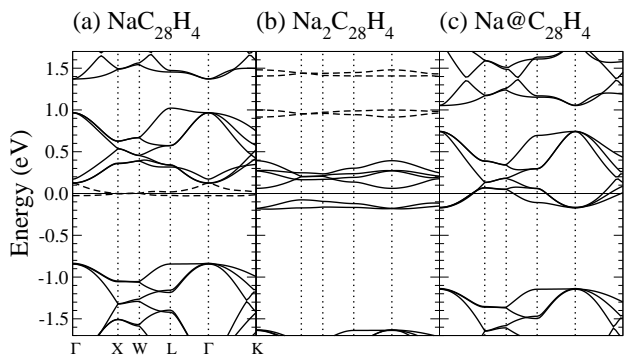}
\caption{\label{doping} Band structure comparison among three doping
cases (a) Intercalation of Na atoms into tetrahedral site; (b)
Intercalation of Na atoms into the interstitial sites between opposing
six-membered rings on nearest neighbor \hsbucky{}; (c) Encapsulation
of Na atom into the \hsbucky{} cage. Dopant bands are depicted by dashed
lines. The thin solid line at zero is the Fermi level.}
\end{figure}

The intercalation of Na atoms into the tetrahedral site occurs with no
significant expansion of the lattice [similar to that found, \eg, in
K$_3$\bucky{} (Ref.~\onlinecite{jlmartins3})]. However, unlike the \bucky{}
fullerides, additional bands appear just below the conduction band of
\hsbucky{} as depicted by the dashed lines in
\Figure~\ref{doping}(a). A Mulliken population analysis reveals that
the H atoms of the \hsbucky{} gain additional charge, which is
reasonable since four H atoms point into each tetrahedral site of
solid \hsbucky{} where a Na atom is located. Hence, in the
Na\hsbucky{} there is no doping of the conduction band; instead the H
and Na states hybridize to form dopant states at the Fermi
level. Furthermore, the calculated enthalpy of reaction indicates that
the compound is not stable, so that this doping scenario would not be
expected to occur naturally.

Intercalation of Na atoms into the interstitial site between opposing
six-membered rings on nearest neighbor molecules results in a doping
ratio twice as high as that of the previous case. Unlike Na\hsbucky{},
there no longer are any dopant bands near the Fermi level.  We
determine the enthalpy of reaction to be -0.9 eV per Na atom which is
in line with those calculated for K$_3$\bucky{}.\cite{jlmartins3} When
the cell stress is relaxed, the lattice constant of Na$_2$\hsbucky{}
increases from 16.3~\AA~to 18.1~\AA{} resulting in a narrowing of the
bands. In \Figure~\ref{doping}(b), we see the opening of a 0.2-eV gap
at the Fermi level and so the solid transforms from a conductor into
an insulator with a small gap upon cell relaxation.  This is a direct
result of Jahn-Teller distortion of the doped \hsbucky{} molecules
whose symmetry is reduced from $T_d$ to $D_{2d}$.  The Jahn-Teller
distortion of the isolated charged \hsbucky{} molecule splits the
3-fold degenerate LUMO into a twofold degenerate and onefold degenerate
state with a 0.2-eV gap.  Hence, when the forces-and-stresses are
relaxed in Na$_2$\hsbucky{}, a band gap similar in size to the
Jahn-Teller gap opens up at the Fermi level. Thus in this case, doping
by intercalation leads to insulating behavior.

The encapsulation of Na atom into the \hsbucky{} cage is the doping
scenario which most closely parallels that of the alkali-doped
\bucky{} solids. Figure~\ref{doping}(c) depicts a simple rigid shift
of the Fermi level into the conduction band of \hsbucky{}. In contrast
to Na$_2$\hsbucky{}, the lattice constant of Na@\hsbucky{} does not
increase.  The Na atom is located at the center of the \hsbucky{} and
we find no Jahn-Teller distortion. Although the isolated charged
\hsbucky{} molecule undergoes a Jahn-Teller distortion due to the
degeneracy of the LUMO state, the broadening of the bands in
Na@\hsbucky{} is found to be sufficient to eliminate this
effect. Since it is known that endohedrally-doped \sbucky{} can be
formed, \cite{tguo1} the relevant enthalpy of reaction for this system
is the total energy of solid Na@\hsbucky{} relative to the isolated
molecules; we find that this solid can be formed with an enthalpy of
reaction similar to that of undoped \hsbucky{}.\footnote{When isolated
molecules are the reference system, the enthalpy of reaction differs
from the binding energy by only a minus sign.}

Within the electron-phonon mediated theory of superconductivity,
$\lambda=N(0)V_\mathrm{ep}$ plays a crucial role in determining
\Tc{}. The superconducting transition temperature for Na@\hsbucky{}
can be estimated using McMillan's solution of the Eliashberg equations,
\cite{wcmcmillan,gmeliashberg}
\begin{eqnarray}
T_c = \frac{\omega_\mathrm{ln}}{1.2}\exp\left[-\frac{1.04(1+\lambda)}{\lambda-\mu^\ast(1+0.62\lambda)}\right] \label{mcmillan}
\end{eqnarray}
where $\omega_\mathrm{ln}$ is a typical phonon frequency and
$\mu^\ast$ is the Coulomb pseudopotential which describes the
effective electron-electron repulsion. Typical values of
$\omega_\mathrm{ln}\approx\,10^3$~K for \hsbucky{} and \bucky{}. We
may expect that $\mu^\ast$ for \hsbucky{} should not differ
considerably from that of \bucky{} since the subbands and phonon
energies are similar. For alkali-doped \bucky{} solids, experimental
results lead to $\mu^\ast\approx\,0.22$.\cite{msfuhrer} The endohedral
doping scenario for \hsbucky{} gives that
$N(0)\approx\,5$~states/eV/spin/molecule which is half of that found
in the canonical alkali-doped \bucky{} structure, \ie,
$\mathrm{K}_3\mathrm{C}_{60}$. \cite{jlmartins3} Accounting for
screening of the $A_g$ phonons in the electron-phonon interaction, a
$V_\mathrm{ep}=169$~meV gives an enhancement factor of
$\lambda(\mathrm{Na}@\mathrm{C}_{28}\mathrm{H}_4)\approx
1.5\lambda(\mathrm{K}_3\mathrm{C}_{60})$.~\footnote{The enhancement
factor was determined using $V_\mathrm{ep}=63$~meV for \bucky{} from
C\^{o}t\'{e} \etal{} (Ref.~\onlinecite{mcote}) and using the calculated
values of $N(0)$ for $\mathrm{K}_3\mathrm{C}_{60}$ and
$\mathrm{Na}@\mathrm{C}_{28}\mathrm{H}_4$, which are in the ratio of
2:1, as discussed in the text.}  These approximations are in line with
those found in the literature \cite{mcote,nbreda,ogunnarsson1} and is
sufficient for qualitative comparison.  Using this value of $\mu^\ast$
and $\omega_\mathrm{ln}$, one finds that
$\lambda(\mathrm{K}_3\mathrm{C}_{60})=0.84$ in order to explain the
experimentally observed
$T_c(\mathrm{K}_3\mathrm{C}_{60})=19.3$~K.\cite{kholczer} The
enhancement factor for $\lambda$ that we calculate predicts a
$T_c(\mathrm{Na}@\mathrm{C}_{28}\mathrm{H}_4) \approx 3
T_c(\mathrm{K}_3\mathrm{C}_{60}) \approx 58$~K, higher than that found
in the highest temperature alkali-doped fulleride [40~K in
Cs$_3$\bucky{} (Ref.~\onlinecite{ttmpalstra})].

In summary, \abinitio{} pseudopotential density functional calculation
were performed on \sbucky{}-derived molecular solids. We find a
\hsbucky{} solid which binds weakly and exhibits many of the band
structure features as \bucky{}.  Several doping scenarios are
investigated and we find that Na@\hsbucky{} to be the most promising
for superconductivity. The calculated electron-phonon interaction
potential and DOS leads to a \Tc{} enhancement of three times that
found in K$_3$\bucky{}. Since endohedral doping is expected to be
insensitive to the type of dopant, this suggests promising
possibilities with other atoms including ones of higher valence that
can control the doping level. Guo \etal{} \cite{tguo1} have succeeded
in encapsulating group-IVB atoms in \sbucky{}, which suggests that
other small fullerenes such as \hsbucky{} can also be endohedrally
doped.

We are grateful to J. Gale, P. Ordej\'{o}n, E. Koch, L. Mitas,
M. C\^{o}t\'{e}, and J. L. Martins for useful discussions. We are also
very thankful to K. Delaney for providing critical readings of this
manuscript. This work was supported by National Science Foundation
Grant No.\ DMR-9976550 and by the U.S.\ Department of Energy under
Contract No.\ DEFG-96-ER45439. This work utilized the NCSA IBM pSeries
690, the PSC HP Alphaserver cluster and the MCC IBM RS/6000 cluster.

\bibliography{sources}

\end{document}